\documentclass[rnote]{aa} % for the research notes
\usepackage{graphicx}
\usepackage[varg]{txfonts}
\usepackage{longtable,lscape}
\begin{document}
%
%   \title{UKIDSS detections of cool brown dwarfs - proper motions, new candidates, and 
%          spectroscopic confirmations\thanks{based on observations with the Large Binocular Telescope (LBT)}}

   \title{UKIDSS detections of cool brown dwarfs - proper motions of 14 known $>$T5 dwarfs and discovery of three new
          T5.5-T6 dwarfs\thanks{based on observations with the Large Binocular Telescope (LBT)}}
%   \subtitle{}

\titlerunning{UKIDSS detections of cool brown dwarfs}

   \author{R.-D. Scholz
          \and
          G. Bihain
          \and
          O. Schnurr
          \and
          J. Storm
          }

   \institute{Leibniz-Institut f\"ur Astrophysik Potsdam (AIP),
              An der Sternwarte 16, 14482 Potsdam, Germany\\
              \email{rdscholz@aip.de, gbihain@aip.de, 
                     oschnurr@aip.de, jstorm@aip.de}
             }

   \date{Received 2 February 2012; accepted 8 April 2012}

% \abstract{}{}{}{}{} 
% 5 {} token are mandatory
 
  \abstract
   {}
   {We contribute to improving the census of cool brown dwarfs (late-T and Y dwarfs)
    in the immediate solar neighbourhood.}
   {By combining near-infrared (NIR) data of UKIDSS with mid-infrared WISE and 
    other available NIR (2MASS) and red optical (SDSS $z$-band) multi-epoch 
    data we detect high proper motion (HPM) objects with colours typical of late 
    spectral types ($>$T5). We use NIR low-resolution spectroscopy for 
    the classification of new candidates.}
   {We determined new proper motions for 14 known 
    T5.5-Y0 dwarfs, many of them being significantly ($>$2-10 times) more
    accurate than previous ones. We 
    detected three new candidates, 
    ULAS~J0954$+$0623, ULAS~J1152$+$0359, and ULAS~J1204$-$0150,
    by their HPMs and colours. Using previously published
    and new UKIDSS positions of the known nearby T8 dwarf WISE~J0254$+$0223 we
    improved its trigonometric parallax to 165$\pm$20~mas. 
    For the three new objects we obtained NIR spectroscopic follow-up 
    with LBT/LUCIFER classifying them as T5.5 and T6 dwarfs. With their estimated
    spectroscopic distances of about 25-30~pc, their proper motions of about 
    430-650 mas/yr lead to tangential velocities of about 50-80~km/s typical of 
    the Galactic thin disk population.
    }
   {}


   \maketitle
%
%________________________________________________________________
\section{Introduction} \label{Sect_Intro}

Although brown dwarfs (BDs) might be even more numerous than M-type red
dwarf stars, to date we know $\sim$10$\times$ less               
substellar objects than stars in the immediate solar neighbourhood (within $\sim$6~pc).
The BDs are difficult to detect, especially in the optical, due to their
relatively small fluxes and red optical-to-NIR colours. This is 
reinforced by (i) their ``failed star'' nature, as they dim and cool down 
with age, evolving across the M, L, and T spectral types
(Burrows et al.~\cite{burrows01}), and (ii) the typical age of several Gyrs of solar 
neighbourhood objects. Most of the BD neighbours are             
expected to have reached temperatures as low or lower than those of T 
dwarfs ($\sim$500-1300~K), which
is confirmed by the fact that almost 75\% of
the objects with trigonometric parallaxes 
$>$100~mas 
in the compilation of L and T dwarfs of Gelino et al. ~(\cite{gelino11},
hereafter G11) 
are T dwarfs, and their fraction continues to grow, in
particular with late-T ($>$T5) types.

Deep large-area 
optical surveys like the Sloan Digital Sky Survey
data releases (SDSS DR7, Abazajian et al.~\cite{abazajian09};
DR8, 
Aihara et al.~\cite{aihara11})
detected only very nearby late-T dwarfs.  
The Canada-France BD survey is deeper, although smaller in sky area  
(Delorme et al.~\cite{delorme08}, hereafter D08;  
Albert et al.~\cite{albert11}).  New 
near-infrared (NIR) and mid-infrared surveys covering big 
sky portions 
like the Large Area Survey (LAS) within the UKIRT InfraRed Deep Sky Surveys
(UKIDSS)\footnote{The UKIDSS project is defined in Lawrence et al.~(\cite{lawrence07}).
UKIDSS uses the UKIRT Wide Field Camera (WFCAM; Casali et al.~\cite{casali07})
and a photometric system described in Hewett et al.~(\cite{hewett06})
which is situated in the Mauna Kea Observatories (MKO) system (Tokunaga et
al.~\cite{tokunaga02}).
The pipeline processing and science archive are described
in Hambly et al.~(\cite{hambly08}) and Irwin et al.~(\cite{irwin12}).}
that goes deeper than the Two Micron All Sky Survey
(2MASS; Skrutskie et al.~\cite{skrutskie06}),
and the Wide-field Infrared Survey Explorer
(WISE; Wright et al.~\cite{wright10}) that
extends 
the 2MASS to the mid-infrared
over the whole sky, are very sensitive
to late-T dwarfs and even cooler Y dwarfs, as recently shown by
Kirkpatrick et al.~(\cite{kirkpatrick11}, hereafter K11)
and Cushing et al.~(\cite{cushing11}, herafter C11).

In continuation of the study of Scholz~(\cite{scholz10}, hereafter S10),
who used UKIDSS DR6 and SDSS DR7
to detect late-T dwarf candidates, and
Scholz et al.~(\cite{scholz11}, hereafter S11),
who combined the WISE preliminary data release (PDR) with SDSS
(DR7$+$DR8) and 2MASS discovering two very nearby ultracool BDs and
classifying one of them (WISE~J1741$+$2553) spectroscopically as a T9.5 dwarf,
we 
look for further late-T dwarf 
candidates with HPMs. We combine recent
UKIDSS data releases (DR7$+$DR8$+$DR9) 
with WISE, 2MASS, and SDSS, where available. 
The UKIDSS releases represent a growing database, whereas 
SDSS DR7 and DR8 data 
are independent and may differ for a given object.
Sects.~\ref{Sect_PMold} and \ref{Sect_PMnew} provide proper motions
and photometry of 14 previously known and three 
newly found objects, respectively. Sect.~\ref{Sect_plx}
provides 
the parallax measurement of a very nearby object.
Sect.~\ref{Sect_Spec} presents spectroscopic
follow-up for our new candidates, and
Sect.~\ref{Sect_Concl} contains our brief discussion and
conclusions.

%
%________________________________________________________________
\section{Selection of candidates and cross-identification} \label{Sect_Sel}

Nearby cool BDs are expected to be very faint 
HPM objects.
We identify them not only based on their 
characteristic colours in one deep survey
but as moving objects showing up in various surveys 
with time baselines of several years.
For our search of nearby $>$T5 dwarfs, primarily based on
latest UKIDSS, i.e. DR9 LAS and Galactic Clusters Survey (GCS),
and SDSS data, we 
used the same criteria as described in S10:\\

$J > 11$,~~~~~~~ $J$$-$$K < 0$,~~~~~~~ $z$$-$$J > +2.5$.\\

Only late-T dwarfs and possibly low-metallicity T dwarfs
meet these constraints. White dwarfs, late-M
and L dwarfs are excluded by the two colour cuts.
To identify a HPM object, we needed at least two different epochs from
UKIDSS or other available survey data, if it had no suitable SDSS
counterpart.
For additional survey data we looked at 2MASS and other catalogues accessible 
via the CDS\footnote{http://vizier.u-strasbg.fr/cgi-bin/VizieR}.
We also checked 2MASS FITS images 
from IRSA Finder Chart\footnote{http://irsa.ipac.caltech.edu/applications/FinderChart/} 
and detected objects visually 
using ''pick object'' within the ESO skycat tool.
After the WISE PDR, we checked our UKIDSS
candidates, including those with lacking/uncertain proper motions, 
for WISE counterparts. 
In the WISE PDR, we also selected suspected 
HPM objects (no 2MASS 
counterpart within 3~arcsec) with 
late-T dwarf mid-infrared 
colours\footnote{$w1$, $w2$, and $w3$ bands are centered at 3.4~$\mu$m, 4.6~$\mu$m, and 12~$\mu$m}
(similar to the criteria used by S11):\\

$w1$$-$$w2 > +1.5$,~~~~~~$w2$$-$$w3 < +2.5$,~~~~~~$w2 < 15$\\

\noindent and subsequently looked for UKIDSS counterparts.
However, using 
WISE as a starting point, we found only
already known objects. We checked the UKIDSS DR9 LAS, GCS, and 
DR7 Galactic Plane Survey (GPS) for possible measurements of 
the $>$T5 dwarfs discovered by K11, 
to improve their proper motions by involving additional epochs.
UKIDSS counterpart data were found in the
LAS and GCS but not in the GPS.
In Sects.~\ref{Sect_PMold} and~\ref{Sect_PMnew},
we show our proper motion results for objects with
at least three distinct epochs available.

%
%________________________________________________________________
\section{Proper motions \& photometry of known objects} \label{Sect_PMold}

On top of Table~\ref{tab1_photpmSpT}\footnote{Table~\ref{tab1_photpmSpT}
is available electronically only}
we list photometry and our proper motions of
previously known late-T ($>$T5)
dwarfs, which we newly identified in UKIDSS LAS data.
Individual multi-epoch
positions from UKIDSS LAS (and other surveys) are listed in
Tables~\ref{tab2_known_posep}, \ref{tab3_astmgcs}, and \ref{tab4_astm}\footnote{Tables~\ref{tab2_known_posep},
\ref{tab3_astmgcs}, and \ref{tab4_astm} are available electronically only},
where our visual 2MASS detections are marked.
Imaging SDSS and UKIDSS LAS data are available
in all fields/bands of the ten objects. For objects not detected
in any of the bands listed in Table~\ref{tab1_photpmSpT},
we give lower magnitude limits
corresponding to the SDSS depths
(Abazajian et al.~\cite{abazajian09}) and in specific fields of
UKIDSS LAS and GCS (using SQL queries from Dye et al.~\cite{dye06}).
Most of the LAS detections are recent WISE discoveries
(K11).
Full designations of all known objects described
in this paper can be found in
K11
and G11.

We also tried to improve the proper motions for three known $>$T5
dwarfs from Table~6 of S10
(not shown here again) by involving
WISE positions if available. However, this led to larger uncertainties
in $\mu_{\delta}$. In particular, we note that one of these three
(the T7 dwarf SDSS~J1507$+$1027;
Chiu et al.~\cite{chiu06}) is detected by WISE as a red object
($w1$$=$16.30$\pm$0.09, $w2$$=$14.12$\pm$0.05, $w3$$=$12.13$\pm$0.33)
but shifted in declination by about 1.5~arcsec with respect to its
expected position according to the HPM measurement of
S10. According to the WISE PDR
documentation, sources fainter than $w1$$\sim$14.5
may have ''declinations being offset between 0.2 and 1.0~arcsec from the true position''.
This object is not included in K11.
The T6.5 dwarf SDSS~J1346$-$0031 (Tsvetanov et al.~\cite{tsvetanov00};
Burgasser et al.~\cite{burgasser06}) in Table~\ref{tab1_photpmSpT}
is also a red WISE source
($w1$$=$15.41$\pm$0.05, $w2$$=$13.56$\pm$0.04, $w3$$=$11.99$\pm$0.32)
but was not listed by K11 in
their Table~1 (with WISE and NIR photometry of known T dwarfs).

We determined the new proper motions in Table~\ref{tab1_photpmSpT}
from weighted linear fitting over all available multi-epoch positions, including
2MASS and SDSS (assuming 100~mas errors), WISE PDR (with their given errors), and
UKIDSS positions (assuming 70~mas errors) listed
in Tables~\ref{tab2_known_posep}, \ref{tab3_astmgcs} and \ref{tab4_astm}, as well as all
WISE and Spitzer measurements (with their errors) given in K11
if available.
In Sect.~\ref{Sect_plx}, we compare our error assumptions with others.
For faint UKIDSS ($>$18~mag) objects we assumed larger errors (140~mas),
and for faint SDSS ($z$$>$20) objects and our visual 2MASS detections, we
increased the assumed errors to 150~mas and 200~mas, respectively.
For CFBDS~J0059$-$0114 we included the CFBDS position (assuming
150~mas errors) from
D08 in our proper motion fitting.
In case of WISE~J0254$+$0223, we combined 2MASS and SDSS DR8
positions already used by S11 with
Pan-STARRS1 measurements 
from 
Liu et al.~(\cite{liu11}, hereafter L11),
multiple WISE (instead of WISE PDR) and Spitzer epochs provided by
K11, and UKIDSS  DR9 data.
For the latter object, we 
improved its
trigonometric parallax (see Sect.~\ref{Sect_plx}).

Our proper motions of the first six WISE discoveries in
Table~\ref{tab1_photpmSpT} are
at least 2-5 times
more accurate than
previous values in S11 and
K11, whereas those of
CFBDS~J0059$-$0114, ULAS~J0827$-$0204 and
SDSS~J1346$-$0031 agree with more accurate 
other data.
Pinfield et al.~(\cite{pinfield12}) measured the proper motion
of WISE~J0750$+$2725
using only two UKIDSS epochs. 
By involving
the WISE and Spitzer epochs, we obtained a 
smaller
$\mu_{\alpha}\cos{\delta}$, whereas our
$\mu_{\delta}$, despite a poor fit, is in  good agreement with the former value.
For SDSS~J1110$+$0116 
our $\sim$2$\times$ more accurate result agrees well
with that of 
Jameson et al.~(\cite{jameson08}),
who did not use the SDSS.

As WISE~J0750$+$2725 was not detected in  SDSS $z$-, UKIDSS $HK$-bands,
and 2MASS, the maximum epoch difference between
the first epoch UKIDSS $J$-band and the last epoch Spitzer data
was less than four years. The resulting proper motion errors
are relatively large, but the total proper motion is highly significant.
Even larger errors for CFBDS~J0059$-$0114, WISE~J2226$+$0440 and WISE~J2344$+$1034
are due to maximum epoch differences of only about 1.5 to 2 years.
The worst case is WISE~J1311$+$0122, where the UKIDSS
data do not significantly extend the time baseline of only
$\sim$1 year, resulting in the largest proper motion errors.
However, all large proper motion components,
already indicated in K11, were confirmed.

For four of the $>$T5 discoveries of
K11, we found UKIDSS GCS $K$-band data
(Table~\ref{tab1_photpmSpT}). However, 
one of the Y0 dwarfs was not detected therein.
For the others we used the positions of their apparent counterparts
(Table~\ref{tab3_astmgcs})
together with 2010/2011 WISE and Spitzer epochs given in
K11 for new proper motion solutions.
For the T6 dwarf WISE~J0410$+$1411, we included
available SDSS DR8 data as well. We also found 
the known late-T dwarf 2MASS~J0516$-$0445
in our search combining WISE and UKIDSS GCS data.
We achieved $\sim$3$\times$ to $>$10$\times$
more accurate proper motions for the
T dwarfs with GCS detection
(Table~\ref{tab1_photpmSpT}) but not
for the Y0 dwarf WISE~J1541$-$2250, which
lies in a relatively crowded region close to Upper Sco
(Lodieu et al.~\cite{lodieu08}) and has highly discrepant
preliminary trigonometric (2-4~pc)
and photometric ($\sim$8~pc) distances in K11.
If the latter is correct, we would not expect to see a counterpart in the
GCS $K$-band, similar to what we found for the other Y0 dwarf,
WISE~J0410$+$1502, which has according to K11
a photometric distance of $\sim$9~pc. 
Therefore, we think the
GCS counterpart of WISE~J1541$-$2250 and its revised proper motion are
doubtful.

% Table 1
\onllongtabL{1}{
\begin{landscape}
\begin{longtable}{lrcrrrrrrcrr}
\caption{\label{tab1_photpmSpT} SDSS DR7 (DR8) $z$ and UKIDSS DR9 $YJHK$ photometry and old and newly determined proper motions (in mas/yr) of known and new $>$T5 dwarfs}\\
\hline
\hline
                  &            &                &                &       &         &      & \multicolumn{3}{c}{previously measured proper motion} &  \multicolumn{2}{c}{new proper motion (this work)} \\
object            & SpT (Ref)  &   $z$          &  $Y$           & $J$   & $H$     & $K$  &  $\mu_{\alpha}\cos{\delta}$ & $\mu_{\delta}$ & (Ref) & $\mu_{\alpha}\cos{\delta}$ & $\mu_{\delta}$ \\
\hline
known (LAS):\\
CFBDS~J0059$-$0114& T8.5 (1,3) & $\gtrsim$20.5  & 18.77$\pm$0.06 & 18.03$\pm$0.05 & 18.86$\pm$0.25 & 18.40$\pm$0.26                            & $+$879$\pm$008 &  $+$51$\pm$005 & (11) & $+$904$\pm$069 &  $+$22$\pm$128 \\
WISE~J0254$+$0223 & T8.0 (2,3) & 19.86$\pm$0.07$^{\blacklozenge}$ & 17.00$\pm$0.01 & 15.92$\pm$0.01 & 16.29$\pm$0.03 & 16.73$\pm$0.05          &$+$2496$\pm$046 & $+$276$\pm$047 & (2)  &$+$2537$\pm$022 & $+$236$\pm$008 \\
WISE~J0750$+$2725 & T8.5 (3,3) & $\gtrsim$20.5  & 19.75$\pm$0.09 & 18.72$\pm$0.04$^*$ & $\gtrsim$18.7  & $\gtrsim$18.1                         & $-$732$\pm$017 & $-$194$\pm$017 & (12) & $-$666$\pm$037 & $-$202$\pm$076 \\
ULAS~J0827$-$0204 & T5.5 (4,4) & 20.41$\pm$0.17 & 18.36$\pm$0.05 & 17.16$\pm$0.02 & 17.43$\pm$0.05 & 17.52$\pm$0.11                            &  $+$27$\pm$003 & $-$109$\pm$002 & (11) &  $+$24$\pm$012 &  $-$98$\pm$006 \\
WISE~J0929$+$0409$^{\blacktriangleright}$ & T6.5 (3,3) & 20.75$\pm$0.12$^*$& 18.00$\pm$0.02 & 16.87$\pm$0.01 & 17.37$\pm$0.07 & 17.40$\pm$0.09 & $+$105$\pm$374 & $-$773$\pm$367 & (3)  & $+$525$\pm$022 & $-$460$\pm$021 \\
SDSS~J1110$+$0116 & T5.5 (5,7) & 19.68$\pm$0.11 & 17.34$\pm$0.01 & 16.16$\pm$0.01 & 16.20$\pm$0.02 & 16.05$\pm$0.03                            & $-$243$\pm$021 & $-$238$\pm$018 & (13) & $-$237$\pm$008 & $-$266$\pm$003 \\
WISE~J1311$+$0122$^{\blacktriangleright}$ & T9:  (3,3) & $\gtrsim$20.5  & 19.89$\pm$0.10 & 18.97$\pm$0.08 & $\gtrsim$18.8  & $\gtrsim$18.3     &  $-$68$\pm$324 & $-$827$\pm$325 & (3)  & $-$150$\pm$229 & $-$847$\pm$047 \\
SDSS~J1346$-$0031 & T6.5 (6,7) & 19.33$\pm$0.04$^*$& 16.80$\pm$0.01 & 15.64$\pm$0.01 & 15.97$\pm$0.01 & 15.96$\pm$0.02                         & $-$503$\pm$003 & $-$114$\pm$001 & (14) & $-$480$\pm$019 & $-$131$\pm$014 \\
WISE~J2226$+$0440$^{\blacktriangleright}$ & T8/T8.5 (3,3) & $\gtrsim$20.5  & 18.04$\pm$0.03 & 16.90$\pm$0.02 & 17.45$\pm$0.07 & 17.24$\pm$0.09 &  $-$56$\pm$584 & $-$529$\pm$596 & (3)  & $-$150$\pm$082 & $-$516$\pm$128 \\
WISE~J2344$+$1034$^{\blacktriangleright}$ & T9 (3,3) & $\gtrsim$20.5 & 19.88$\pm$0.12 & 18.84$\pm$0.09 & 19.24$\pm$0.29 & $\gtrsim$18.2        & $+$574$\pm$667 & $-$211$\pm$668 & (3)  & $+$859$\pm$114 &  $-$10$\pm$128 \\
\hline
known (GCS):\\
WISE~J0307$+$2904 & T6.5 (3,3)& n/a                           & - & n/a & n/a            & 18.08$\pm$0.12                &  $+$91$\pm$307 &  $-$79$\pm$312 & (3)  & $-$154$\pm$025 &  $-$23$\pm$016 \\
WISE~J0410$+$1502 & Y0 (3,8)  & $\gtrsim$20.5                 & - & n/a & n/a            & $\gtrsim$18.2  &  \\
WISE~J0410$+$1411 & T6 (3,3)  & 20.52$\pm$0.12$^{\blacklozenge}$ & - & n/a & n/a            & 17.82$\pm$0.20             &  $+$19$\pm$302 & $-$198$\pm$326 & (3)  &  $-$81$\pm$037 &  $-$65$\pm$047 \\
2MASS~J0516$-$0445& T5.5 (9,7)& 19.31$\pm$0.08$^{\blacklozenge}$ & - & n/a & 15.77$\pm$0.01 & 15.79$\pm$0.02             & $-$270$\pm$030 & $-$210$\pm$030 & (15) & $-$225$\pm$007 & $-$193$\pm$009 \\
WISE~J1541$-$2250$^{??}$ & Y0 (3,8)  & n/a                           & - & n/a & n/a            & 17.93$\pm$0.19         & $-$780$\pm$234 & $-$218$\pm$249 & (3)  & $+$984$\pm$293 & $+$516$\pm$187 \\
\hline
new (LAS):\\
ULAS~J0954$+$0623 & T5.5  (10,10) & 19.58$\pm$0.11 & 17.87$\pm$0.02 & 16.59$\pm$0.01 & 16.89$\pm$0.02 & 17.17$\pm$0.07   &                &                &      & $-$495$\pm$014 & $-$422$\pm$008 \\
ULAS~J1152$+$0359 & T6    (10,10) & 20.69$\pm$0.19 & 18.54$\pm$0.03 & 17.28$\pm$0.02 & 17.70$\pm$0.05 & 17.77$\pm$0.12   &                &                &      & $-$388$\pm$017 & $-$209$\pm$033 \\
ULAS~J1204$-$0150 & T5.5  (10,10) & 20.49$\pm$0.20 & 17.99$\pm$0.03 & 16.81$\pm$0.02 & 17.07$\pm$0.04 & 17.29$\pm$0.09   &                &                &      & $-$399$\pm$016 & $+$166$\pm$011 \\
\hline
\end{longtable}

\smallskip

\footnotesize{\textbf{Notes:} $^{\blacktriangleright}$ not in WISE PDR,
$^{\blacklozenge}$ SDSS DR8,
$^*$ mean values from multiple measurements,
$^{??}$  doubtful GCS counterpart/new proper motion,
n/a - not available,
$z$ magnitudes in AB system,
$YJHK$ {\it{= aperMag3}} for point sources
(Dye et al.~\cite{dye06}) in
Vega system using MKO photometric system.
References (discovery, SpT, proper motions) are:
(1) - D08,
(2) - S11,
(3) - K11,
(4) - Lodieu et al.~(\cite{lodieu07}),
(5) - Geballe et al.~(\cite{geballe02}),
(6) - Tsvetanov et al.~(\cite{tsvetanov00}),
(7) - Burgasser et al.~(\cite{burgasser06}),
(8) - C11,
(9) - Burgasser et al.~(\cite{burgasser03}),
(10) - New (this paper, SpT
accurate to $\pm$0.5 sub-types),
(11) - Marocco et al.~(\cite{marocco10}),
(12) - Pinfield et al.~(\cite{pinfield12}),
(13) - Jameson et al.~(\cite{jameson08}),
(14) - Tinney, Burgasser \& Kirkpatrick~(\cite{tinney03}),
(15) - Faherty et al.~(\cite{faherty09}).}
\end{landscape}
}% End onllongtabL

% Table 2 available electronically only
\onltab{2}{
\begin{table}
\caption{$\alpha, \delta$ (J2000.0)
of known $>$T5 dwarfs in UKIDSS LAS, etc.}
\label{tab2_known_posep}
\centering
\begin{tabular}{llcrl}
\hline\hline
Object \\
$\alpha$ [deg] & $\delta$ [deg] & Epoch & ID & Source \\
\hline
\multicolumn{5}{l }{CFBDS~J0059$-$0114$^{1}$:} \\
 14.7949709 & $-$01.2336947 & 2005.732 &  292702 & ULAS $H$ \\
 14.7949181 & $-$01.2337060 & 2005.732 &  293033 & ULAS $K$ \\
 14.7952080 & $-$01.2337464 & 2006.885 & 1581752 & ULAS $Y$ \\
 14.7952146 & $-$01.2337521 & 2006.885 & 1584033 & ULAS $J$ \\
\hline
\multicolumn{5}{l }{WISE~J0254$+$0223$^{2,3}$:} \\
 43.5399358 & $+$02.3996412 & 2010.734 & 3859643 & ULAS $H$ \\
 43.5399337 & $+$02.3996466 & 2010.734 & 3859715 & ULAS $K$ \\
 43.5399295 & $+$02.3996464 & 2010.718 & 3840324 & ULAS $Y$ \\
 43.5399246 & $+$02.3996529 & 2010.718 & 3840484 & ULAS $J$ \\
\hline
\multicolumn{5}{l }{WISE~J0750$+$2725$^{3}$:} \\
117.5164404 & $+$27.4293472 & 2007.126 & 1231545 & ULAS $J$ \\
117.5159819 & $+$27.4292434 & 2009.129 & 2367125 & ULAS $Y$ \\
117.5159827 & $+$27.4292398 & 2009.129 & 2367307 & ULAS $J$ \\
\hline
\multicolumn{5}{l }{ULAS~J0827$-$0204$^{4}$:} \\
126.7819292 & $-$02.0687722 & 1998.953 &         & 2M $J^{\star}$ \\
126.781892  & $-$02.068816  & 2001.293 &    2259 & SDSS \\
126.7819523 & $-$02.0689431 & 2005.940 &  465119 & ULAS $Y$ \\
126.7819511 & $-$02.0689491 & 2005.940 &  473125 & ULAS $J$ \\
126.7819424 & $-$02.0689509 & 2005.940 &  478440 & ULAS $H$ \\
126.7819506 & $-$02.0689568 & 2005.940 &  477757 & ULAS $K$ \\
\hline
\multicolumn{5}{l }{WISE~J0929$+$0409$^{3}$:} \\
142.276914  & $+$04.167209  & 2001.140 &    2125 & SDSS \\
142.2776989 & $+$04.1663975 & 2007.107 & 1219425 & ULAS $K$ \\
142.277670  & $+$04.166407  & 2007.197 &    6749 & SDSS \\
142.2777383 & $+$04.1663741 & 2007.288 & 1399129 & ULAS $H$ \\
142.2781424 & $+$04.1660432 & 2009.923 & 3028681 & ULAS $J$ \\
142.2781455 & $+$04.1660397 & 2009.923 & 3029035 & ULAS $Y$ \\
\hline
\multicolumn{5}{l }{SDSS~J1110$+$0116:} \\
167.541711  & $+$01.270302  & 2000.126 &         & 2M \\
167.541674  & $+$01.270298  & 2000.343 &    1462 & SDSS \\
167.5411353 & $+$01.2697012 & 2008.373 & 2325845 & ULAS $H$ \\
167.5411422 & $+$01.2697045 & 2008.373 & 2325865 & ULAS $K$ \\
167.5410610 & $+$01.2695783 & 2009.978 & 3085143 & ULAS $Y$ \\
167.5410582 & $+$01.2695767 & 2009.978 & 3085325 & ULAS $J$ \\
\hline
\hline
\multicolumn{5}{l }{WISE~J1311$+$0122$^{3}$:} \\
197.7760412 & $+$01.3817322 & 2010.156 & 3351901 & ULAS $Y$ \\
197.7760299 & $+$01.3817386 & 2010.156 & 3352083 & ULAS $J$ \\
\hline
\multicolumn{5}{l }{SDSS~J1346$-$0031:} \\
206.693495  & $-$00.530639  & 1999.221 &     756 & SDSS \\
206.693104  & $-$00.530592  & 2001.093 &         & 2M \\
206.692471  & $-$00.530844  & 2006.395 &    6166 & SDSS \\
206.6921079 & $-$00.5309516 & 2009.129 & 2369313 & ULAS $H$ \\
206.6921118 & $-$00.5309485 & 2009.129 & 2369436 & ULAS $K$ \\
206.6921010 & $-$00.5309524 & 2009.134 & 2350848 & ULAS $Y$ \\
206.6921028 & $-$00.5309471 & 2009.134 & 2351038 & ULAS $J$ \\
206.6920776 & $-$0.5309553  & 2010.182 &         & WISE \\
\hline
\multicolumn{5}{l }{WISE~J2226$+$0440$^{3}$:} \\
336.5960939 & $+$04.6678106 & 2009.620 & 2784734 & ULAS $Y$ \\
336.5961029 & $+$04.6678081 & 2009.620 & 2783676 & ULAS $J$ \\
336.5960903 & $+$04.6678034 & 2009.641 & 2797400 & ULAS $H$ \\
336.5960866 & $+$04.6677827 & 2009.641 & 2797524 & ULAS $K$ \\
\hline
\multicolumn{5}{l }{WISE~J2344$+$1034$^{3}$:} \\
356.1923904 & $+$10.5710685 & 2009.668 & 2823367 & ULAS $Y$ \\
356.1924237 & $+$10.5710277 & 2009.668 & 2823532 & ULAS $J$ \\
356.1924059 & $+$10.5710051 & 2009.827 & 2944838 & ULAS $H$ \\
\hline
\end{tabular}
%
%\smallskip
%
%\scriptsize{
%Notes:
\tablefoot{
$^{1}$ - additional epoch in D08,
$^{2}$ - more data in
S11 (SDSS DR8, 2MASS),
L11 (Pan-STARRS1),
$^{3}$ - WISE and Spitzer epochs
in K11,
$^{4}$ - seen in WISE $w2$ image but no catalogue entry,
$^{\star}$ - our visual detection.
IDs are run and multiframe numbers
for SDSS and UKIDSS LAS (ULAS), respectively.
}
\end{table}
}% end of onltab

% Table 3 available electronically only
\onltab{3}{
\begin{table}
\caption{$\alpha, \delta$ (J2000.0) of known $>$T5 dwarfs in UKIDSS GCS, etc.}
\label{tab3_astmgcs}
\centering
\begin{tabular}{lllrl}
\hline\hline
Object \\
$\alpha$ [deg] & $\delta$ [deg] & Epoch & ID & Source \\
\hline
\multicolumn{5}{l }{WISE~J0307$+$2904$^{1}$:} \\
46.8525439 & $+$29.0798528 & 2008.786 & 2334900 & UGCS $K$ \\
\hline
\multicolumn{5}{l }{WISE~J0410$+$1411$^{1}$:} \\
62.7270953 & $+$14.1920928 & 2005.781 &  322532 & UGCS $K$ \\
62.7269979 & $+$14.1921757 & 2006.367 &    6003 & SDSS \\
\hline
\multicolumn{5}{l }{2MASS~J0516$-$0445:} \\
79.039410  & $-$04.763872  & 1998.721 &         & 2M       \\
79.0392237 & $-$04.7640431 & 2000.934 &    1927 & SDSS \\
79.0386926 & $-$04.7644903 & 2009.893 & 2998751 & UGCS $H$ \\
79.0386834 & $-$04.7644923 & 2009.893 & 2998937 & UGCS $K$ \\
79.0386734 & $-$04.7645655 & 2010.182 &         & WISE \\
\hline
\multicolumn{5}{l }{WISE~J1541$-$2250$^{1}$:} \\
235.4642206& $-$22.8407768 & 2008.301 & 2323120 & UGCS $K$$^{??}$ \\
\hline
\end{tabular}
%
%\smallskip
%
%\scriptsize{
%Notes:
\tablefoot{
$^{1}$ - additional WISE and Spitzer epochs
in K11,
$^{??}$ - doubtful counterpart.
ID are run and multiframe numbers
for SDSS DR8 and UKIDSS GCS (UGCS), respectively.
}
\end{table}
}% end of onltab

% Table 4 available electronically only
\onltab{4}{
\begin{table}
\caption{Multi-epoch $\alpha, \delta$ (J2000.0) of new late-T dwarf candidates}
\label{tab4_astm}
\centering
\begin{tabular}{lllrl}
\hline\hline
Object \\
$\alpha$ [deg] & $\delta$ [deg] & Epoch & ID & Source \\
\hline
\multicolumn{5}{l }{ULAS~J0954$+$0623:} \\
148.625883  &  $+$06.387122 & 2000.153 &         & 2M $J$$^{\star}$ \\
148.625654  &  $+$06.386935 & 2002.174 &    3015 & SDSS \\
148.6249784 & $+$06.3863407 & 2007.214 & 1287023 & ULAS $K$ \\
148.6249374 & $+$06.3863301 & 2007.277 & 1390233 & ULAS $H$ \\
148.6246477 & $+$06.3860939 & 2009.332 & 2442500 & ULAS $J$ \\
148.6245730 & $+$06.3860068 & 2009.942 & 3052277 & ULAS $Y$ \\
\hline
\multicolumn{5}{l }{ULAS~J1152$+$0359:} \\
178.124725  &  $+$03.991431 & 2000.181 &         & 2M $J$$^{\star}$ \\
178.124063  &  $+$03.991187 & 2006.023 &    5973 & SDSS \\
178.1238153 & $+$03.9910013 & 2008.375 & 2326168 & ULAS $H$ \\
178.1237922 & $+$03.9910487 & 2008.375 & 2326188 & ULAS $K$ \\
178.1236578 & $+$03.9909038 & 2009.978 & 3084746 & ULAS $Y$ \\
178.1236518 & $+$03.9908986 & 2009.978 & 3084839 & ULAS $J$ \\
\hline
\multicolumn{5}{l }{ULAS~J1204$-$0150:} \\
181.187079  &  $-$01.843475 & 1999.077 &         & 2M $J$$^{\star}$ \\
181.187242  &  $-$01.843383 & 1999.077 &         & 2M $H$$^{\star}$ \\
181.187080  &  $-$01.843431 & 2000.116 &    1140 & SDSS \\
181.1861442 & $-$01.8430247 & 2008.395 & 2327790 & ULAS $H$ \\
181.1861488 & $-$01.8430134 & 2008.395 & 2327810 & ULAS $K$ \\
181.1861335 & $-$01.8430218 & 2008.403 & 2329084 & ULAS $Y$ \\
181.1861363 & $-$01.8430238 & 2008.403 & 2329105 & ULAS $J$ \\
\hline
\end{tabular}
%
%\smallskip
%
%\scriptsize{
%Notes:
\tablefoot{
ID are run and multiframe numbers
for SDSS and UKIDSS LAS (ULAS), respectively.
$^{\star}$ indicate our visual detections.
}
\end{table}
}% end of onltab

%_____________________________________________________________
%                 A figure as large as the width of the column
%-------------------------------------------------------------
   \begin{figure}
   \centering
   \includegraphics[width=86mm,angle=0]{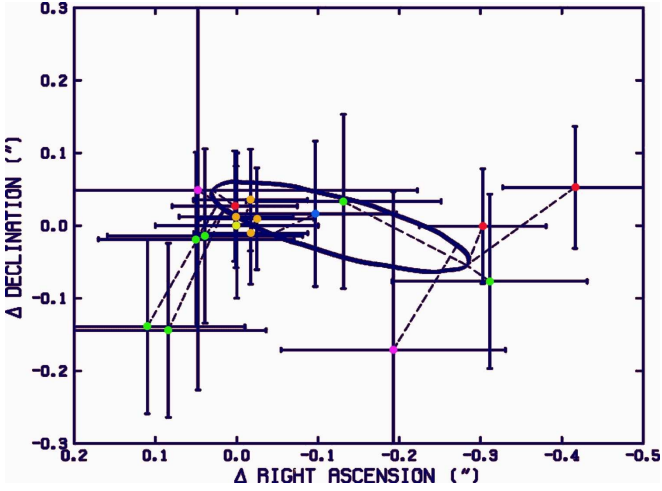}
      \caption{Parallactic ellipse for WISE~J0254$+$0223
               after combined proper motion and parallax solution
               and subtraction of the proper motion.
               Coloured (blue - SDSS, green -
               Pan-STARRS1, yellow - 2MASS, orange - UKIDSS, red - WISE,
               pink - Spitzer) dots with error bars show
               the individual positional measurements, dashed
               lines show their displacements relative to the fit.
              }
         \label{fig1_plx}
   \end{figure}

%
%________________________________________________________________
\section{Trigonometric parallax for WISE~J0254$+$0223} \label{Sect_plx}

L11 and K11 published
preliminary trigonometric parallaxes
for WISE~J0254$+$0223, which has the largest proper motion 
in Table~\ref{tab1_photpmSpT}. By combining all available
measurements used by them (less than 10 epochs in both cases)
with new UKIDSS DR9 data
(Table~\ref{tab2_known_posep}), we collected
17 epochs and expected an improved fit.
We applied the software of Gudehus~(\cite{gudehus01})
for combined proper motion and parallax solutions with and without
weights corresponding to the positional errors of the different surveys. 
In our final weighted solution, which yielded $\sim$10\% more accurate
parallax and proper motion results compared to the unweighted one,
we used the following errors: 70~mas for UKIDSS (as we can assume 
for an object at intermediate galactic latitude according to 
Lawrence et al.~\cite{lawrence07}), 100~mas for 2MASS and SDSS, 120~mas
for Pan-STARRS1 (corresponding to the typical size of the error
bars shown in Fig.~4 of L11, and the individual
errors of the three WISE (between 76~mas and 89~mas) and two Spitzer
epochs (from 138~mas to 275~mas) given in 
K11. Our assumed 2MASS errors
are smaller than the values given in K11
and correspond to those of L11. For the
SDSS counterpart we assumed slightly larger errors than 
L11 as we took into account that the 
astrometric accuracy at the faint end of the SDSS
is limited by photon statistics (Pier et al.~\cite{pier03}).
Our proper motion and parallax of WISE~J0254$+$0223 are:\\

$(\mu_{\alpha}\cos{\delta}$, $\mu_{\delta})$$=$($+$2544$\pm$7, $+$237$\pm$7)~mas/yr,~~ $\pi$$=$165$\pm$20~mas.\\
 
The proper motion is more accurate
than the one in Table~\ref{tab1_photpmSpT}  
from linear fitting not taking into account the 
parallactic motion. In Fig.~\ref{fig1_plx} 
we show the parallactic ellipse and the individual positions
with their error bars and with respect to their expected location 
(indicated by dashed lines) according to the fit 
after subtracting the proper motion.
Only in four cases the shift between expected location
and measured position exceeds the error bars. This concerns
three out of six Pan-STARRS1 and one out of three WISE positions,
whose errors may be underestimated.
 
The distance estimate of 6.1$\pm$0.7~pc from the new parallax
agrees well with the photometric distances of 
5.5$^{+2.3}_{-1.6}$~pc (S11),
7.2$\pm$0.7~pc (L11), and
$\sim$6.9~pc (K11).
Our new parallax is also very similar to previous 
trigonometric measurements
of 171$\pm$45~mas (from 9 positions) and 165$\pm$46~mas (from
6 positions), respectively by L11 and
K11, 
but has a 50\% smaller error. However, our 
new parallax is still a preliminary result, not obtained
within a dedicated parallax programme and making use of 
different optical, NIR, and mid-infrared surveys. 
As we neglected colour-dependent systematic errors
in the positions (differential refraction), 
our parallax errors are lower limits.
We also derived significant parallaxes for
ULAS~J0827$-$0204 (97$\pm$21~mas), SDSS~J1110$+$0116 (56$\pm$5~mas), and
the new T5.5 dwarf (Sects.~\ref{Sect_PMnew} and \ref{Sect_Spec}) ULAS~J0954$+$0623 (136$\pm$28~mas), but these results are
uncertain (from only six epochs in each case).

%_____________________________________________________________
%                 A figure as large as the width of the column
%-------------------------------------------------------------
   \begin{figure}
   \centering
   \includegraphics[width=56mm,angle=270]{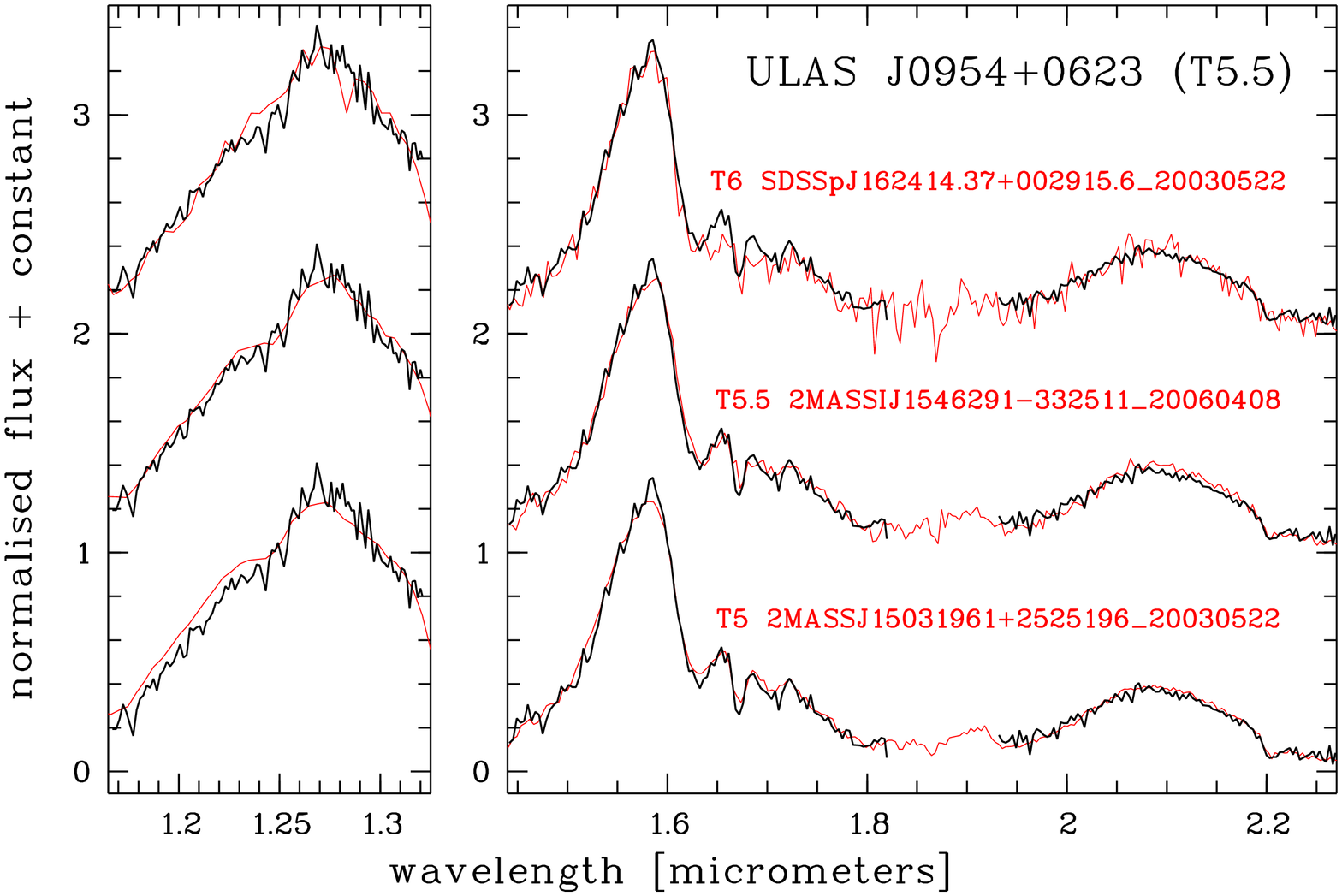}
   \includegraphics[width=56mm,angle=270]{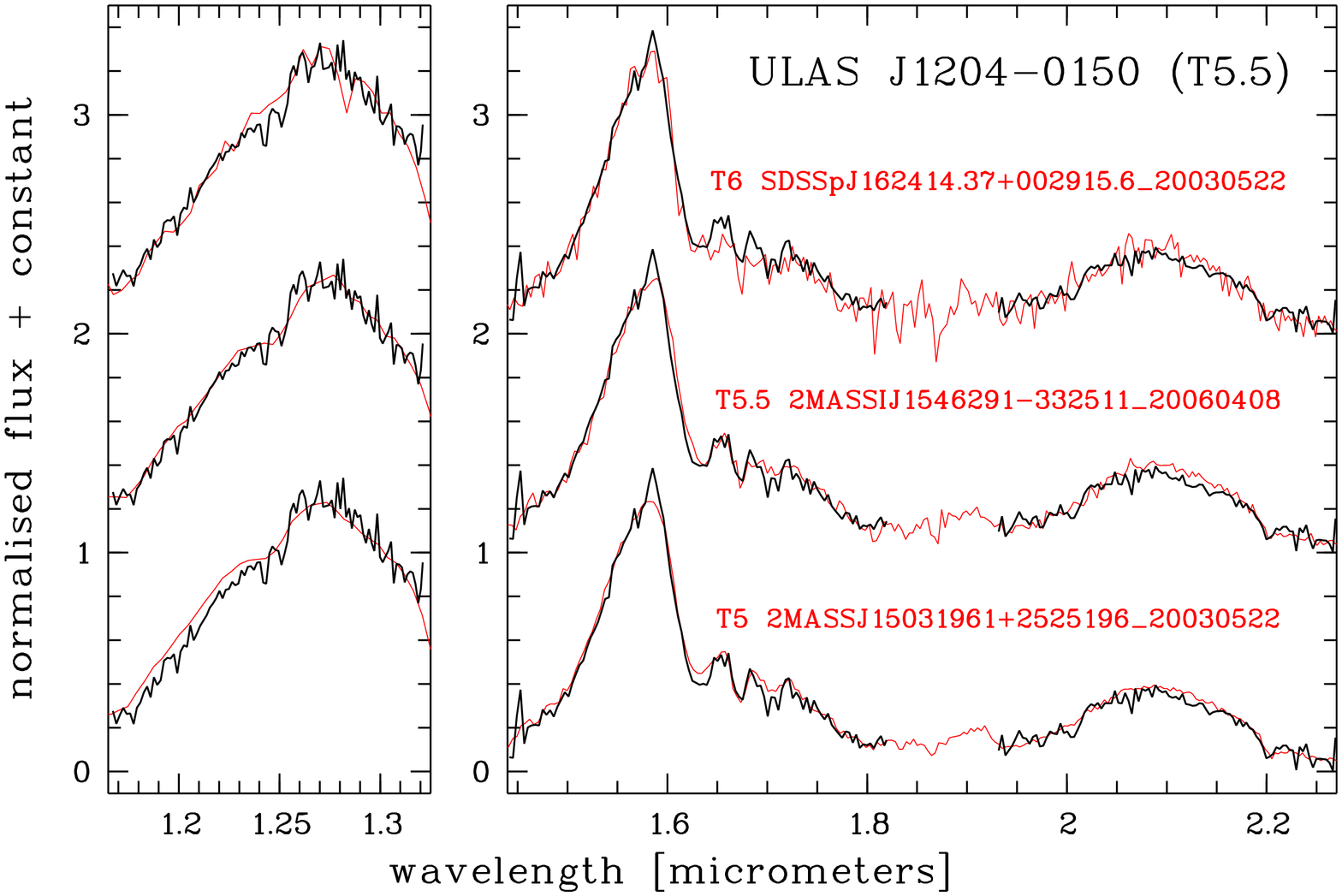}
   \includegraphics[width=56mm,angle=270]{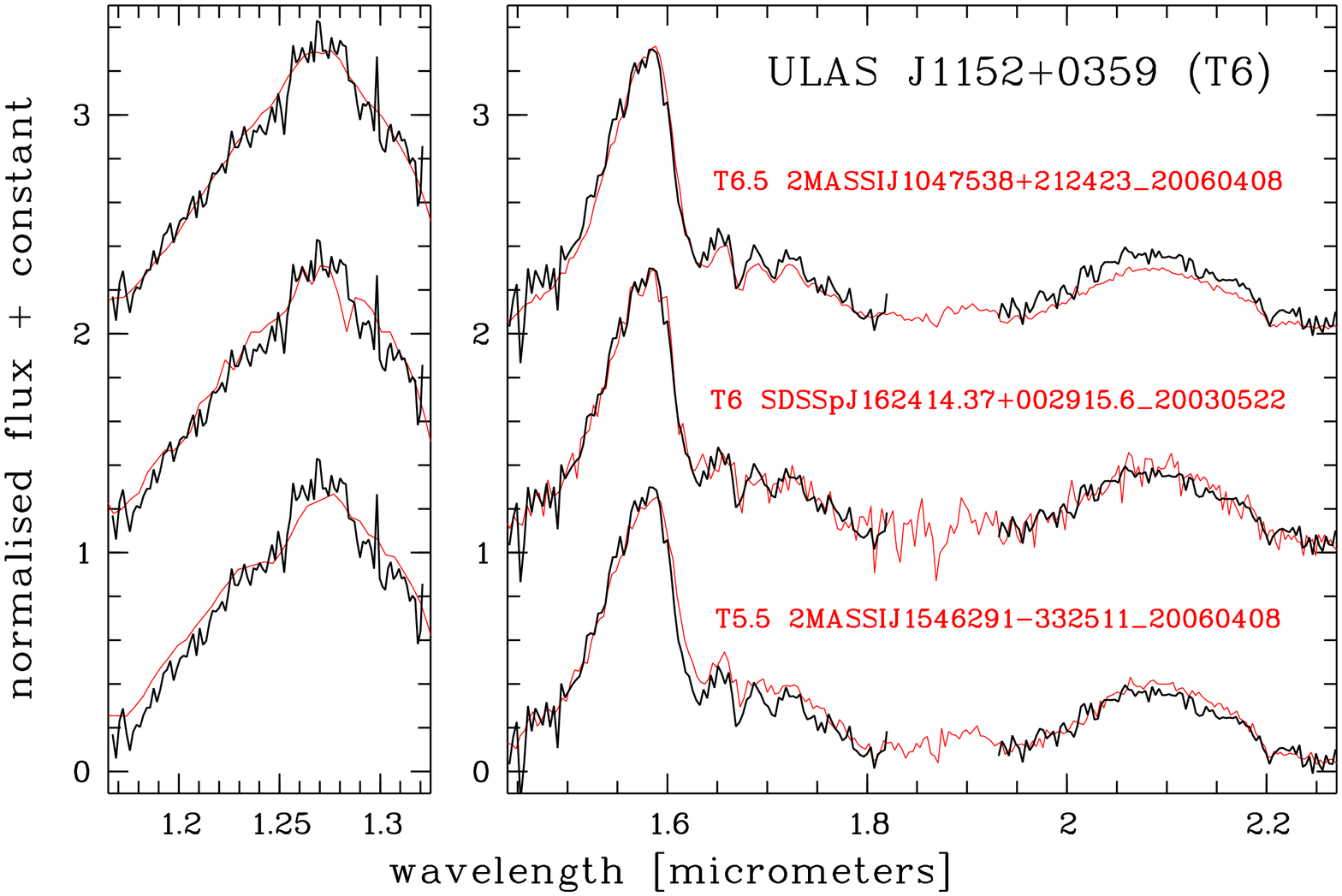}
      \caption{LBT/LUCIFER1 $J$- (left) and $H$$+$$K$-band
               (right) spectra of
               ULAS~J0954$+$0623 (top), ULAS~J1204$-$0150 (centre), and
               ULAS~J1152$+$0359 (bottom). Target spectra (black) were
               smoothed for the comparison with the lower-resolution ($R$$\sim$120)
               template spectra (red) of T5-T6.5 dwarfs 
               from Adam Burgasser's 
               SpeX Prism Spectral Libraries.
               Original references for the template spectra are:
               T5, T6 - Burgasser et al.~(\cite{burgasser04});
               T5.5, T6.5 - Burgasser et al.~(\cite{burgasser08}).}
         \label{fig2_spec}
   \end{figure}

%
%________________________________________________________________
\section{New late-T dwarf candidates} \label{Sect_PMnew}

Using the UKIDSS LAS as a starting point,
we discovered three new late-T dwarf candidates:
ULAS~J095429.91$+$062309.9,
ULAS~J115229.67$+$035927.2, 
ULAS~J120444.67$-$015034.8 (hereafter names are abbreviated).
They have $z$-only
SDSS counterparts 
shifted by several arcsec with respect to their 
two LAS epochs.
For our proper motion fits we measured their
positions at a fourth (earliest) epoch by visually detecting 2MASS
counterparts (Table~\ref{tab4_astm}). The resulting well-measured
HPMs, and the $zYJHK$ magnitudes of the objects 
are shown in Table~\ref{tab1_photpmSpT}. WISE PDR data are not available
for these objects.

All three new candidates can be preliminarily classified
as T7$\pm$2 dwarfs based on their negative
NIR colours $-$0.41$<$$J$$-$$H$$<$$-$0.26 and $-$0.58$<$$J$$-$$K$$<$$-$0.48.
As shown by S10, this colour-based classification cannot
be made more accurately for this range of negative NIR colour indices.
This is due to the colour reversal of T8-T9 dwarfs, whereas the sequence of
T dwarfs with earlier types shows a clear trend with both colours, which allows
a more accurate colour-based classification for positive and moderately negative
indices (see Fig.~3 in S10).
The more accurate spectral types (SpT) given in Table~\ref{tab1_photpmSpT}
are the result of our spectroscopic follow-up described in the next
section.

% Table 5 available electronically only
\onltab{5}{
\begin{table*}
\caption{Spectral indices and corresponding types and classification from comparison with templates for three new late-T dwarfs}
\label{tab5_spind}
\centering
\begin{tabular}{llll}
\hline\hline
Index & ULAS~J0954$+$0623 & ULAS~J1152$+$0359 & ULAS~J1204$-$0150 \\
\hline
H$_2$O-H & 0.319$\pm$0.012 (T6/T5) & 0.299$\pm$0.021 (T6) & 0.315$\pm$0.015 (T6/T5) \\
CH$_4$-H & 0.347$\pm$0.012 (T6/T5) & 0.304$\pm$0.016 (T6) & 0.369$\pm$0.012 (T5/T6) \\
CH$_4$-K & 0.233$\pm$0.022 (T5/T4) & 0.199$\pm$0.030 (T5) & 0.225$\pm$0.024 (T5) \\
SpT(J)   & T5.5                  & T6.5               & T5.5  \\
SpT(HK)  & T5.5                  & T6                 & T5.5  \\
SpT$_{adopted}$ & T5.5           & T6                 & T5.5  \\
\hline
\end{tabular}
\end{table*}
}% end of onltab

%
%________________________________________________________________
\section{Spectroscopic observations with LBT/LUCIFER} \label{Sect_Spec}

For follow-up we used the
Large Binocular Telescope (LBT) NIR spectrograph LUCIFER1
(Mandel et al.~\cite{mandel08}; Seifert et al.~\cite{seifert10};
Ageorges et al.~\cite{ageorges10}) in long-slit
spectroscopic mode with
the $H$$+$$K$ (200 lines/mm + order separation filter) 
and $zJHK$ gratings (210 lines/mm + $J$ filter). 
ULAS~J1204$-$0150 and ULAS~J0954$+$0623 were observed
with total integrations of 30~min in $H$$+$$K$ and 20~min
in $J$ on 11 April 2011 and 12 April 2011, respectively.
The slighly fainter ULAS~J1152$+$0359 was observed 
on 11 May 2011 with 36~min 
in $H$$+$$K$ and 24~min in $J$.
As in S11, central
wavelengths were chosen at 1.835~$\mu$m ($H$$+$$K$) and
1.25~$\mu$m ($J$) yielding a coverage of 1.38--2.26 and
1.17--1.32~$\mu$m, respectively. The slit width was 1~arcsec 
for ULAS~J1152$+$0359, 
corresponding to 4 pixels, with a spectral resolving power 
$R$=$\lambda$/$\Delta$$\lambda$$\approx$4230, 940, 
and 1290 at $\lambda$$\approx$1.24,
1.65, and 2.2~$\mu$m, respectively. 
For the other objects the slit width was 2~arcsec.
Observations consisted of individual
exposures of 75~s in $H$$+$$K$ and 200~s in $J$ 
(90 and 240~s, respectively, for ULAS J1152+0359) 
with shifting the target along the slit 
using an ABCCBA pattern
until the total integration time was reached.
A0V standards were observed just 
before/after 
the targets with similar airmasses.

The raw spectroscopic data were reduced as described in 
S11
using standard routines in {\tt
IRAF\footnote{IRAF is distributed by the National Optical Astronomy
Observatories, which are operated by the Association of Universities for
Research in Astronomy, Inc., under cooperative agreement with the National
Science Foundation.}}. 
The calibrated $J$ and $H$$+$$K$ spectra of our new objects and the corresponding
spectra of the templates shown in Fig.~\ref{fig2_spec}  were normalised by the
average flux in the range 1.20--1.30 and 1.52--1.61~$\mu$m, respectively.
Each spectrum was classified by visual comparison with template spectra,
separately for the $J$- and $H$$+$$K$-band,
and by three measured spectral indices in the $H$$+$$K$-band 
(Table~\ref{tab5_spind}\footnote{Table~\ref{tab5_spind} is available electronically only}) 
as defined by Burgasser et al.~(\cite{burgasser06}). 
No other useful indices could be determined in the $J$-band, since
the LUCIFER $J$ grating provides a very narrow wavelength interval.
Fig.~\ref{fig2_spec} shows the three closest
matches of template (in red) with smoothed target spectra (overplotted in black). 
Both methods, comparison with templates and 
classification with indices, provided consistent results 
(with $\pm$0.5 sub-types accuracy, as in S11):
ULAS~J0954$+$0623 and ULAS~J1204$-$0150 were classified as T5.5, whereas 
ULAS~J1152$+$0359 turned out to be a T6 dwarf. 

Using the
relationship between spectral types and absolute $JHK$ magnitudes
determined by Marocco et al.~(\cite{marocco10}) from L0-T9 dwarfs with
known trigonometric parallaxes (excluding known and possible binaries),
we get $M_J$$=$14.70, $M_H$$=$14.92, $M_K$$=$14.99 for T5.5 and
$M_J$$=$14.92, $M_H$$=$15.22, $M_K$$=$15.32 for T6 dwarfs.  
Assuming conservatively
absolute magnitude uncertainties
of about $\pm$0.4~mag, involving
our $\pm$0.5 sub-types classification,
we get spectroscopic distances of
25$\pm$5~pc, 31$\pm$6~pc, and 27$\pm$6~pc, respectively
for ULAS~J0954$+$0623, ULAS~J1152$+$0359, and ULAS~J1204$-$0150.

%
%________________________________________________________________
\section{Discussion and conclusions} \label{Sect_Concl}

Searching HPM $>$T5 dwarfs in the 
latest UKIDSS LAS/GCS data
we detected some previously known objects, including
nine recent WISE discoveries (K11), and revised
their proper motions. For the very nearby T8 dwarf 
WISE~J0254$+$0223 (S11, L11,
K11) we derived
an improved trigonometric parallax.
The coolest known objects in our sample, 
two recently classified Y0 dwarfs (C11), 
were either too faint to be detected
or have a doubtful counterpart in the UKIDSS GCS.

We found new late-T candidates in the UKIDSS LAS and confirmed
their $>$T5 types by LBT/LUCIFER1 NIR spectroscopy. 
ULAS~J0954$+$0623, ULAS~J1152$+$0359, and ULAS~J1204$-$0150 are 
classified as T5.5, T6, and T5.5 dwarfs, respectively, all
residing within $\sim$30~pc from the sun.
From their total proper motions of 650, 441, and 432~mas/yr,
we compute tangential velocities of 77$\pm$15, 65$\pm$13, and
55$\pm$12~km/s, respectively. These values are
smaller than the 87$\pm$8~km/s of WISE~J0254$+$0223 
for which L11 found thin disk membership.
We conclude that the new nearby cool BDs
also belong kinematically to 
the thin disk population of the Galaxy.

More accurate proper motions allow a more 
reliable determination of companionship to nearby stars.
We checked the previously known objects (with our new proper motions)
and our newly discovered ones for possible wide companions with common proper
motions in SIMBAD\footnote{http://simbad.u-strasbg.fr} or
in the database on L and T dwarfs provided by G11. 
Using a search radius of one degree around the target positions, we find no 
bright or faint wide companion
candidate for which the proper motion agrees within the error bars and
the projected separation is 
smaller than about 10000 AU,
if our estimated (or known) distances are taken into acount.

With our HPM- 
and colour-based survey, we have added three relatively
nearby objects to the pool of $\sim$40 previous
UKIDSS discoveries of cool ($>$T5) BDs
(Warren et al.~(\cite{warren07}; 
Lodieu et al.~\cite{lodieu07}; 
Chiu et al.~\cite{chiu08};
Pinfield et al.~\cite{pinfield08};
Lodieu et al.~\cite{lodieu09};
Lucas et al.~\cite{lucas10};
Burningham et al.~\cite{burningham10}).
Most of them were identified based
on their colours without proper motion measurements.
Our three new objects are among the five brightest (in $J$)
of all $>$T5 UKIDSS discoveries.

%
%________________________________________________________________
\begin{acknowledgements}
The authors thank Roland Gredel, Jochen Heidt, Jaron Kurk, Ric Davies,
and all observers at the LBT for assistance during the
preparation and execution of LUCIFER observations, and
Adam Burgasser for providing template spectra
at http://pono.ucsd.edu/$\sim$adam/browndwarfs/spexprism.

This research has made use of the WFCAM Science Archive 
providing UKIDSS, the
NASA/IPAC Infrared Science Archive, which is operated by the Jet Propulsion
Laboratory, California Institute of Technology, under contract with the
National Aeronautics and Space Administration,
and of data products from WISE,
which is a joint project of the University of California, 
Los Angeles, and the Jet Propulsion Laboratory/California Institute of 
Technology, funded by the National Aeronautics and Space Administration,
from 2MASS, and from SDSS DR7 and DR8. 
Funding for SDSS-III has been provided by the Alfred P. Sloan Foundation, 
the Participating Institutions, the National Science Foundation, and the 
U.S. Department of Energy. The SDSS-III web site is http://www.sdss3.org/.
We used SIMBAD and VizieR at the CDS/Strasbourg.

\end{acknowledgements}

%
%________________________________________________________________

\end{document}